\documentstyle[prl,aps,epsf,rotate,multicol]{revtex}

\begin{document}
\draft

\title{Large Scale Cross-Correlations in Internet Traffic}

\author{Marc Barth\'elemy$^1$, Bernard Gondran$^2$, and Eric
Guichard$^3$}

\address{
$^1$ CEA, Service de Physique de la Mati\`ere Condens\'ee,
BP12 Bruy\`eres-Le-Ch\^atel, France\\
$^2$ R\'eseau National de T\'el\'ecommunications pour\\ 
la Technologie, l'Enseignement et la Recherche\\
151, Bld de L'H\^opital, 75013 Paris, France\\
$^3$ Equipe R\'eseaux, Savoirs \& Territoires\\
Ecole normale sup\'erieure, 75005 Paris, France
}

\date{\today}
\maketitle
\begin{abstract}

The Internet is a complex network of interconnected routers and the
existence of collective behavior such as congestion suggests that the
correlations between different connections play a crucial role. It is
thus critical to measure and quantify these correlations. We use
methods of random matrix theory (RMT) to analyze the cross-correlation
matrix {\bf \sf C} of information flow changes of $650$ connections
between $26$ routers of the French scientific network `Renater'. We
find that {\bf \sf C} has the universal properties of the Gaussian
orthogonal ensemble of random matrices: The distribution of
eigenvalues---up to a rescaling which exhibits a typical correlation
time of the order $10$ minutes---and the spacing distribution follows
the predictions of RMT. There are some deviations for large
eigenvalues which contain network-specific information and which
identify genuine correlations between connections. The study of the
most correlated connections reveal the existence of `active centers'
which are exchanging information with a large number of routers
thereby inducing correlations between the corresponding
connections. These strong correlations could be a reason for the
observed self-similarity in the WWW traffic.

\end{abstract}



\begin{multicols}{2}

\section{Introduction}

Internet connects different routers and servers using different
operating systems and transport protocols.  This intrinsic
heterogeneity of the network added to the unpredictability of human
practices\cite{Huberman97} make the Internet inherently unreliable and
its traffic
complex\cite{Leland94,Csabai94,Thompson97,Feldmann98,Takayasu00}. Recently,
there has been major advances in our understanding of the generic
aspects of the Internet\cite{IMP,Faloutsos99,Caldarelli00,Pastor01}
and web\cite{Kumar99,Broder00,Albert99,Adamic99,Huberman98,Mossa02}
structure and development, revealing that these networks can exhibit
emergent collective behavior characterized by scaling. Concerning data
transport, most of the studies focus on properties at short time
scales (usually $<1$ min) or at the level of individual
connections\cite{Leland94,Crovella97,Willinger97}. In particular, it
has been shown that for wide- and local-area networks the
self-similarity (for time correlations) applies. Possible reasons for
this behavior were shown to be\cite{Crovella97} the underlying
distribution of WWW documents, the effects of user `think time', and
the addition of many such transfers.

Studies on statistical flow properties at a large scale
\cite{Csabai94,Thompson97,Takayasu00,Fukuda99} concentrate essentially
on the phase transition from a `fluid' regime to a `congested' one for
which the average packet travel time is very large\cite{Ohira98}. The
existence of such a collective behavior indicates the importance of
spatial correlations between connections at a large scale in the
system.  In order to be able to understand and to model the traffic in
the network, it is thus important to measure and to quantify the
correlations between the flows in different connections.

In this paper, we analyze the correlations between different
connections of a wide area network which is the French scientific
network `Renater'. We use random matrix theory (RMT) to study the
corresponding empirical correlation matrix. RMT has been developed in
the fifties for studying complex energy levels of heavy
nuclei\cite{Mehta} and more recently it has also been used in the
study of correlations of stocks\cite{Bouchaud99,Amaral99} or
statistics of atmospheric correlations \cite{Santha01}.

We first demonstrate the validity of the universal predictions of RMT
for the eigenvalue statistics of the cross-correlation
matrix. However, we observe some deviations compared to the minimal
hypothesis of random independent time-series. These deviations from
the universal predictions of RMT identify system-specific, non-random
properties of the network providing clues about the nature of the
underlying interactions. This result allows one to distinguish genuine
correlations in the network which are not just due to noise.

\section{Empirical results}

\subsection{Data studied}

\begin{figure}
\narrowtext
\centerline{
\epsfysize=0.8\columnwidth{{\epsfbox{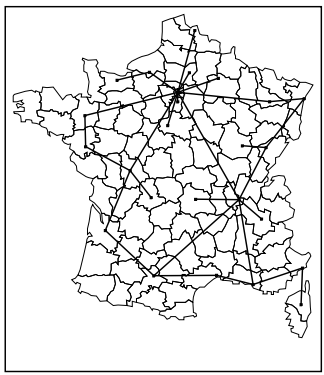}}}
}
\vspace*{0.1cm}
\caption{ Map of the Renater network. There is a total of about $30$
interconnected routers (of which $26$ are effectively studied). We
show on this map the physical connections. The measured data consist
in a flow matrix $F_{ij}(t)$ (with $t=\tau m$, $m=0,\cdots, L-1$ and
$i,j=1,\cdots,26$) which gives the effective flow exchange between
routers $i$ and $j$. For more details on this network, see the web
page {\sf http://www.renater.fr} and for an animated version of flows,
see {\sf http://barthes.ens.fr/metrologie/Renater01}.}
\label{france}
\end{figure}

We use data from the French network `Renater'\cite{Renater} which has
about $2$ million users and which consists of about $30$
interconnected routers (Fig.~\ref{france}). Most Research institutes,
technological, or educational institutions are connected to Renater.

The data consist of the real exchange flow (sum of Ftp, Telnet, Mail,
Web browsing, etc.) between all routers even if there is not a direct
(physical) link between all of them. For a connection $(i,j)$ between
routers $i$ and $j$ ($i\neq j$), $F_{ij}(t)$ (in bytes per $5$
minutes) is the effective information flow at time $t$ going out from
$i$ to $j$ (the flow going from $i$ to $k$ via $j$ is excluded from
$F_{ij}$). For technical reasons, data for a few routers were not
reliable and we analyzed data for $26$ routers which amounts in
$26\times 26$ matrices $F_{ij}(t)$ given for every sampling time scale
$\tau=5$ minutes during a two weeks period. We also exclude from the
present study the internal flow $F_{ii}$, and the nights for which the
flow is essentially due to machine activity. We thus studied data for
days ($8$am-$6$pm), which amounts to a total of $N=26\times 25=650$
different connections given for $L=12\times 10\times 14\/{\rm
days}=1680$ time counts. We choose as a measure of the magnitude of
the time-series fluctuations the growth rate defined as the logarithm
of the ratio of successive counts
\begin{equation}
g_{ij}(t)=\log\left[\frac{F_{ij}(t+\tau)}{F_{ij}(t)}\right]
\end{equation}
for $t=0,\cdots,(L-1)\tau$.  This measure has several nice properties.
First, any multiplicative, time-independent sample bias cancels in the
ratio.  Second, this measure has a natural interpretation in terms of
relative growth since for a small increase $g_{ij}(t)\simeq
[F_{ij}(t+\Delta t)-F_{ij}(t)]/F_{ij}(t)$ is simply the relative
increment. A large value of this quantity reflects a large activity
(i.e. a large flow variation), while a small value corresponds to an
almost constant flow. This measure is thus independent from the volume
of information exchanged and thus does not eliminate the `small'
routers. The study of volume flow exchange will be published
elsewhere\cite{Barthelemy02} and in the present paper the quantity
$g$ allows us to study more subtle effects such as the activity of a
regional router, independently of its `size' measured in terms of
exchanged information volume.

\subsection{Correlation matrix}

The simplest measure of correlations between different connections
$(i,j)$ and $(k,l)$ is the equal-time cross-correlation matrix {\bf \sf
C} which has elements
\begin{equation}
C_{(ij)(kl)} = { \langle g_{ij} g_{kl} \rangle - \langle g_{ij}
\rangle \langle g_{kl} \rangle \over \sigma_{ij} \sigma_{kl} } \,
\label{eq.1}
\end{equation}
where $\sigma_{ij}= \sqrt{\langle g_{ij}^2 \rangle - \langle g_{ij}
\rangle^2} $ is the standard deviation of the flow growth rate of the
connection $(i,j)$ and $\langle\cdots\rangle$ denotes a time average
over the period studied. The correlation matrix is real symmetric and
its elements are comprised between $-1$ (anti-correlated connections)
and $1$ (correlated connections), while a null value denotes
statistical independence. 

The quantities $g_{ij}/\sigma_{ij}$ have (by construction) a variance
equal to one and a zero mean (for a sufficiently long time). It is
thus natural to compare our empirical results with a mutual
independent time-series---the `null' hypothesis---described by the
correlation matrix
\begin{equation}
{\bf \sf R}=\frac{1}{L}{\bf \sf A}{\bf \sf A}^t
\end{equation}
where {\bf \sf A} (the so-called random Wishart matrix) is an $N\times
L$ matrix containing N times series of L random independent elements
with zero mean and unit variance (${\bf \sf A}^t$ denotes the
transpose of ${\bf \sf A}$). Each element of $\sf R$ can be written as
$R_{(ij)(kl)} = \langle a_{ij} a_{kl}\rangle$ where $a_{ij}(t)$ is a
time series of independent elements with zero mean ($\langle
a_{ij}\rangle=0$) and unit variance ($\sigma_{ij}=1$).

\subsubsection{Eigenvalues}

\begin{figure}
\narrowtext
\centerline{
\epsfysize=0.7\columnwidth{{\epsfbox{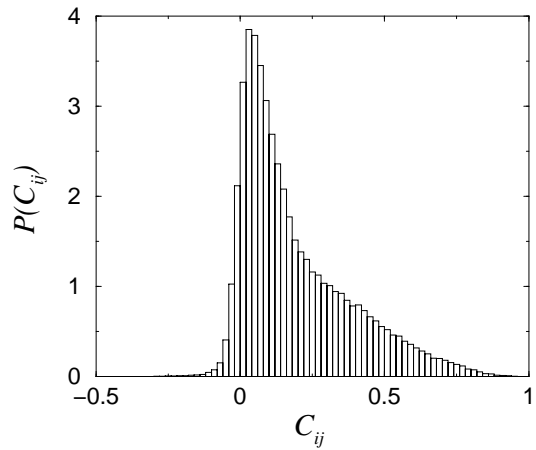}}}
}
\vspace*{0.1cm}
\caption{ Probability distribution for the correlation coefficient
calculated from $5$-minutes flows in the Renater network for a $14$
days period. The average value is positive indicating strong
correlations among the whole network.}
\label{PC}
\end{figure}
\begin{figure}
\narrowtext
\centerline{
\epsfysize=0.5\columnwidth{\rotate[r]{\epsfbox{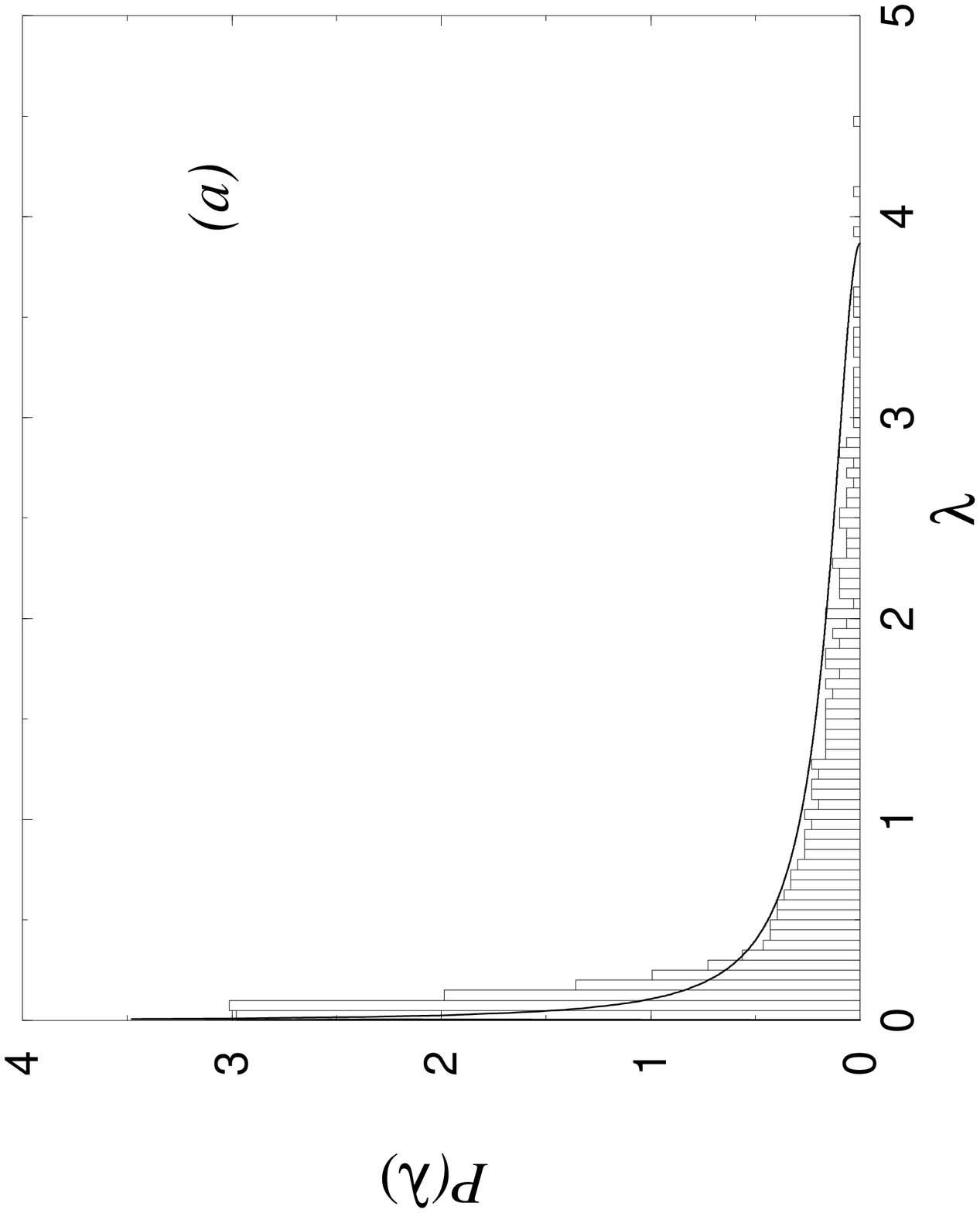}}}
\hspace*{0.1cm}
\epsfysize=0.5\columnwidth{\rotate[r]{\epsfbox{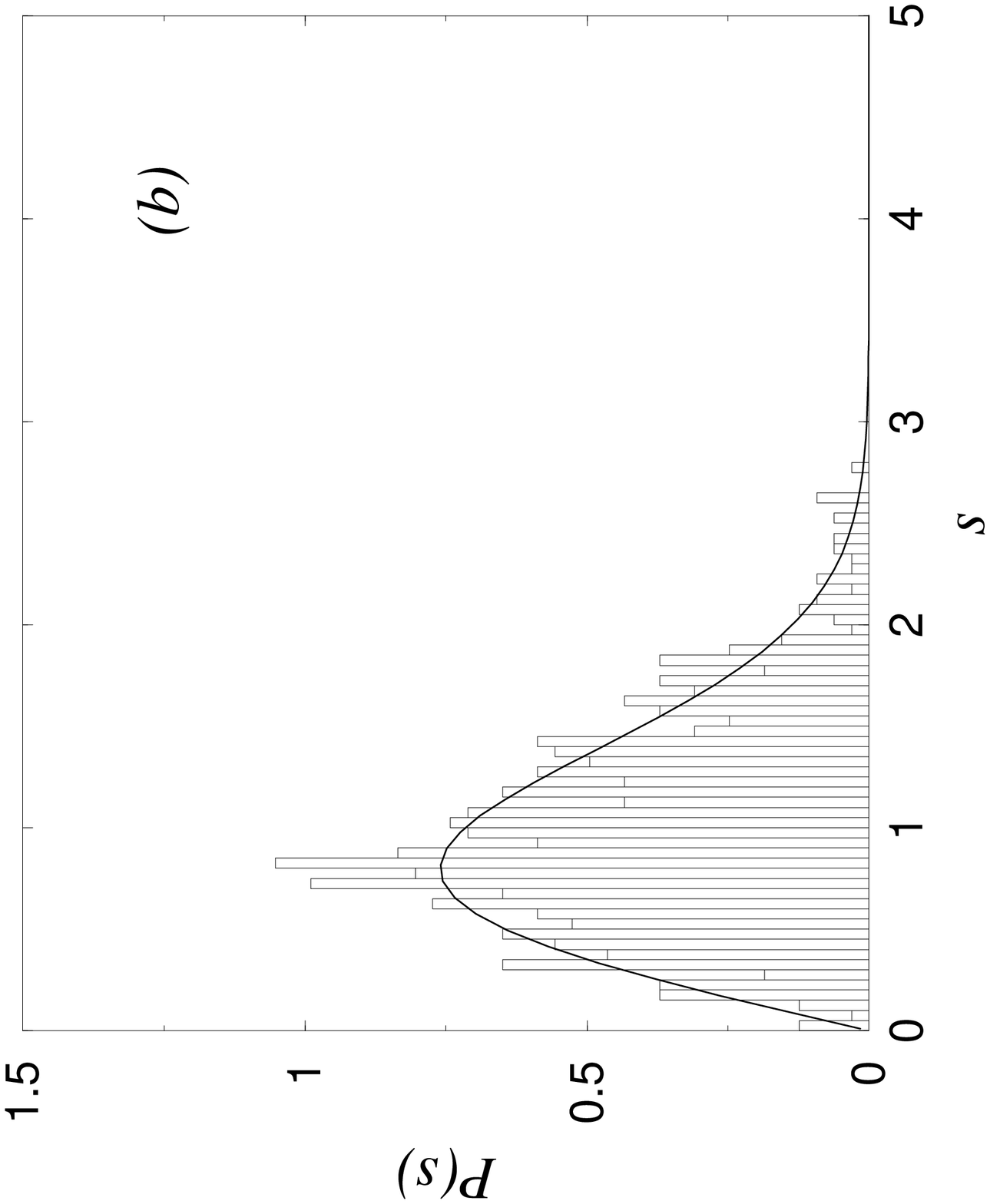}}}}
\vspace*{0.5cm}
\caption{ {\bf (a)} The probability density of the eigenvalues of the
normalized cross-correlation matrix {\bf \sf C} for the 650
connections for a 2-weeks period. The results are reasonably fitted by
the analytical result obtained for cross-correlation matrices
generated from uncorrelated time series (solid line, obtained from
Equ.~$4$ with $Q^*=1.1$). There are however very large eigenvalues
(not shown), the largest one being of order $200$. {\bf (b)}
Nearest-neighbor spacing distribution of the eigenvalues of {\bf \sf
C} after unfolding using the Gaussian broadening
procedure~\protect\cite{unfold}. The solid line is the RMT prediction
for the spacing distribution for the Gaussian orthogonal ensemble
(GOE).}
\label{p_lam}
\end{figure}

The probability distribution of the elements of {\bf \sf C} shows that
most on the elements are positive (Fig.~\ref{PC}) which indicates a
strong correlation among the whole network. For comparison, the
elements of {\bf \sf R} are distributed according to a centered
distribution with zero mean. We now study the statistical properties
of {\bf \sf C} by applying RMT techniques. We first diagonalize {\bf
\sf C} and obtain its eigenvalues $\lambda_k$ ($k=1,\cdots,N$) which
we sort from the largest to the smallest. We then calculate the
eigenvalue distribution and compare it with the analytical result for
a cross-correlation matrix generated from finite uncorrelated time
series~\cite{Edelman88} in the limit $N\to\infty$, $L\to\infty$ where
$Q=L/N\ge 1$ is fixed
\begin{equation}
P_{{\text rm}}(\lambda)=\frac{Q}{2\pi}
\frac{
\sqrt{(\lambda_+-\lambda)(\lambda-\lambda_-)}
}
{\lambda}
\label{p_lamb}
\end{equation}
with $\lambda\in[\lambda_-,\lambda_+]$ and where
\begin{equation}
\lambda_\pm(Q)=1+1/Q\pm 2/\sqrt{Q}
\end{equation}
The eigenvalue distribution of {\bf \sf C} is very different from
Equ.~(\ref{p_lamb}) which predicts a finite range of eigenvalues
depending on the ratio $Q$. The theoretical value is $Q=2.58$ and we
can reasonably fit the empirical curve with an effective value
$Q^*=1.1$ (Fig.~\ref{p_lam}a). This effective value can be explained
as resulting from time correlations in the traffic of the order of
$\frac{Q}{Q^*}\times\tau\simeq 11$ minutes. However, even this fit
cannot reproduce the large eigenvalues observed: For $Q^*=1.1$ the
theoretical eigenvalues are distributed in the interval $2.17\times
10^{-3}\leq \lambda_k \leq 3.82$ while few---a total of order
$20$---measured eigenvalues (not all shown on the graph) are found
above $\lambda_+(Q^*)=3.82$. The largest eigenvalue is of order
$\lambda_1\simeq 200$ namely approximately hundred times larger than
the maximum eigenvalue predicted for uncorrelated time series. As we
will see, the empirical distribution of eigenvector components for the
large eigenvalues is `flat', all components being of the same
order. This suggests that the largest eigenvalues are associated with
strong correlations among the network.


We also calculate the distribution of the nearest-neighbor spacings
$s=\lambda_{k+1} -\lambda_k$. We compare the empirical distribution
of nearest-neighbor spacings with the RMT predictions for real
symmetric random matrices. This class of matrices shares universal
properties with the ensemble of matrices whose elements are
distributed according to a Gaussian probability measure---the Gaussian
orthogonal ensemble (GOE). We find good agreement (Fig.~\ref{p_lam}b)
between the empirical data and Wigner's surmise
\begin{equation}
P_{\rm GOE}(s)= {\pi s \over 2}\, \exp\left(- {\pi \over 4}\, 
s^2 \right)\,.
\label{eq.3}
\end{equation}
which indicates a `level repulsion' existing in our system and means
that the eigenvalues are correlated.

\subsubsection{Eigenvectors and Inverse Participation Ratio}

\begin{figure}
\narrowtext
\centerline{
\epsfysize=0.6\columnwidth{\rotate[r]{\epsfbox{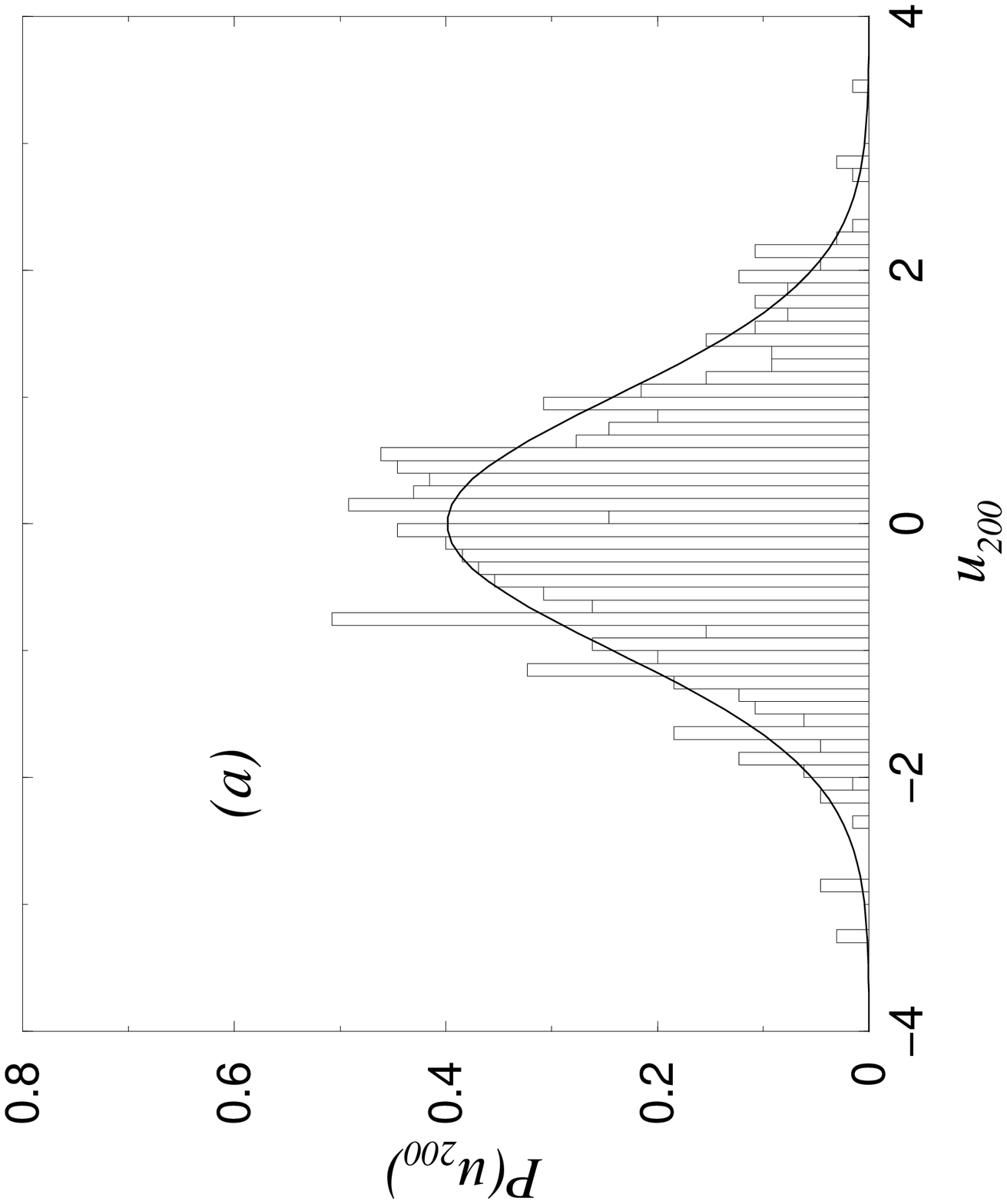}}}
}
\vspace*{0.1cm}
\centerline{
\epsfysize=0.5\columnwidth{\rotate[r]{\epsfbox{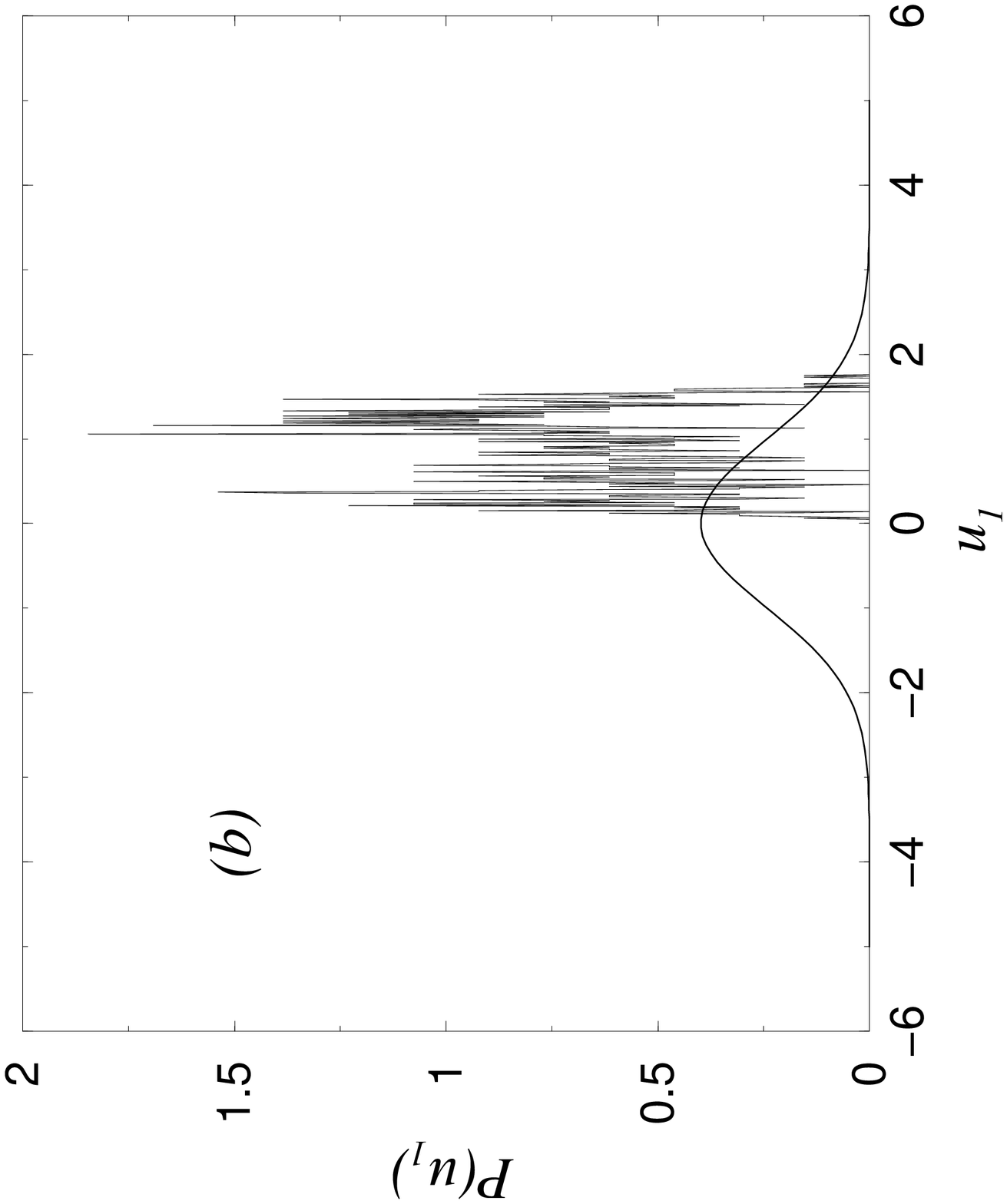}}}
\hspace*{0.1cm}
\epsfysize=0.5\columnwidth{\rotate[r]{\epsfbox{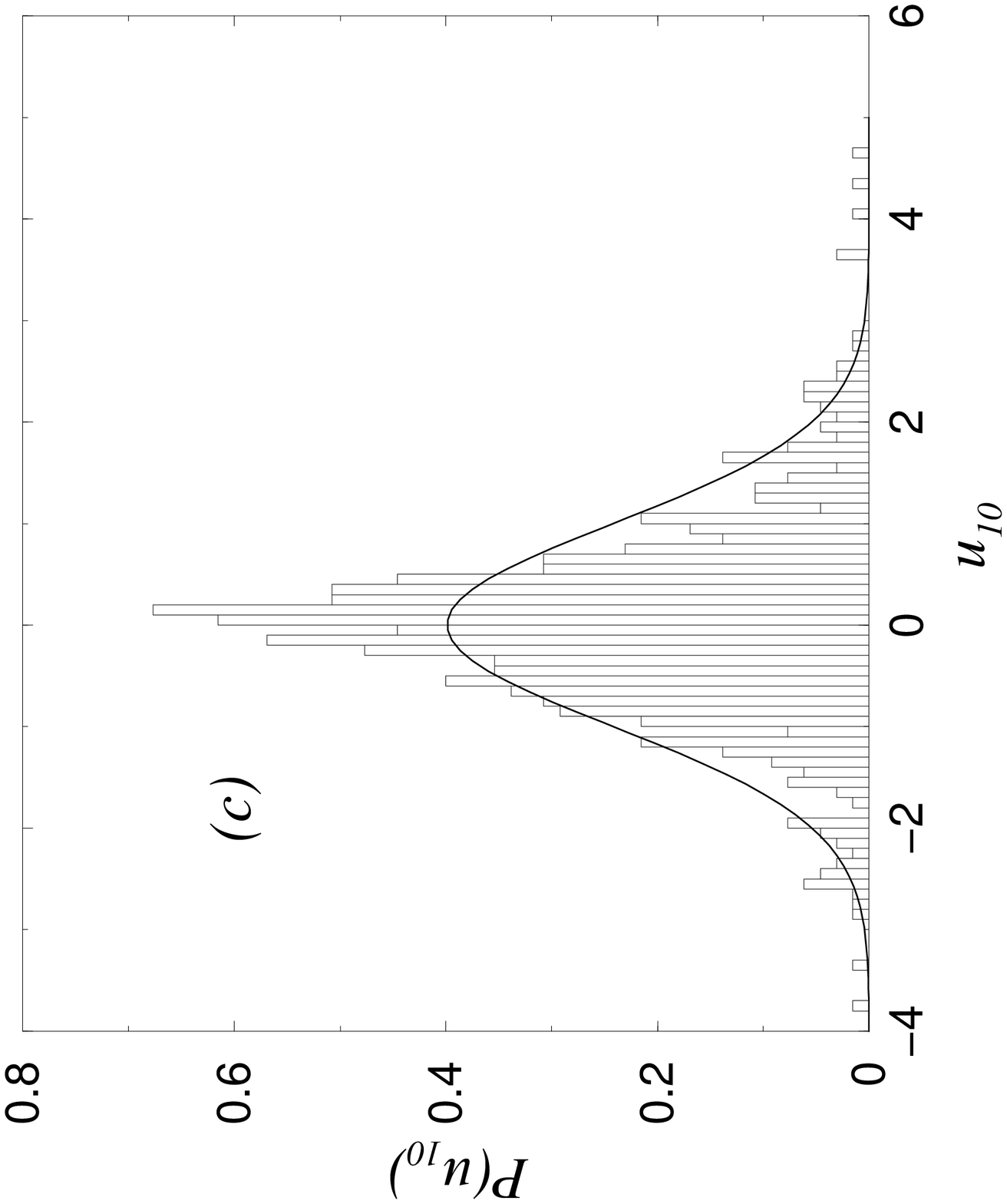}}}
}
\vspace*{0.1cm}
\caption{ Eigenvector component distribution (a) For eigenvalues in
the center of the spectrum. In this case, the empirical results are in
agreement with the results of RMT which is the Porter-Thomas
distribution represented by a solid line. (b,c) For large eigenvalues
there is a clear deviation compared to RMT predictions represented by
the solid line (Porter-Thomas distribution). For the largest
eigenvalue, most of the components is non-zero and positive which
indicates correlations among the whole network.}
\label{p_vector}
\end{figure}

We now analyze the eigenvectors of {\bf \sf C}. We denote by $u_k$ the
eigenvector associated to the eigenvalue $\lambda_k$ and if we
normalize the eigenvectors such that $u_k^2=N$, it can be shown that
in the Wishart case the components $u$ of the eigenvectors are
distributed according to the so-called Porter-Thomas distribution
\begin{equation}
P(u)=\frac{1}{\sqrt{2\pi}}e^{-u^2/2}
\label{Porter}
\end{equation}
In agreement with this result we find that eigenvectors corresponding
to most eigenvalues in the `bulk' of the spectrum ($\lambda_k$ not too
large) follow this prediction (Fig.~\ref{p_vector}a).

On the other hand, eigenvectors with eigenvalues outside the bulk
($\lambda_k\ge\lambda_+(Q^*)$) show marked deviations from the
Gaussian distribution (Fig.~\ref{p_vector}b,c). In particular, the
vector corresponding to the largest eigenvalue $\lambda_1$ deviates
significantly from the Gaussian distribution predicted by RMT
(Fig.~\ref{p_vector}b). This eigenvector is the signature of a
collective behavior---the network itself--- for which all connections
are correlated. This effect was already observed in the framework of
stock correlations, the largest eigenvalue being in this case the
entire market\cite{Bouchaud99,Amaral99,Amaral99b}.

\begin{figure}
\narrowtext
\centerline{
\epsfysize=0.8\columnwidth{\rotate[r]{\epsfbox{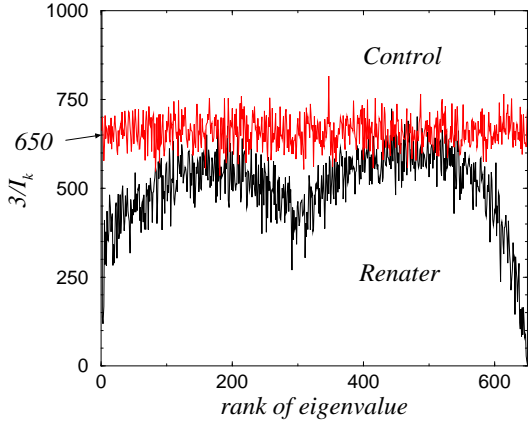}}}
}
\vspace*{0.1cm}
\caption{ Reciprocal inverse participation ratio for each of the $650$
eigenvectors (sorted for decreasing eigenvalues). As a control case,
we show the corresponding result for uncorrelated independent time
series of the same length as the data. Empirical data show small
values at both edges of the spectrum, whereas the control shows only
small fluctuations around the average value $\langle 3/I
\rangle=N=650$.}
\label{ratio}
\end{figure}

The distribution of the components of an eigenvector contains
information about the number of connections contributing to it. In
order to distinguish between one eigenvector with approximately equal
components and another with a small number of large components we use
the inverse participation ratio (IPR) introduced in the context of
localization theory\cite{Wegner80,Revmodphys}
\begin{equation}
I_{k}=\frac{1}{N^2}\sum_{i=1}^N [u_{ki}]^4 \ ,
\label{e.ipr}
\end{equation}
where $u_{ki}$, $i=1,\dots,N=650$ are the components of eigenvector
$u_k$. When the components of a vector are of the same order and
distributed according to Equ.~(\ref{Porter}), the average IPR is small
and equal to $3/N$ whereas a vector with only few non zero components
leads to a IPR of order unity. The quantity $\Upsilon_k=3/I_k$ is thus
a measure of the number of vector components significantly different
from zero. We compared $\Upsilon_k$ for our empirical results and for
uncorrelated time series with the same values of $(N,L)$
(Fig.~\ref{ratio}). For the latter case, $\Upsilon_k$ has small
fluctuations around $N=650$ indicating that all the vectors are
extended\cite{Revmodphys} which means that almost all connections
contribute to them. On the other hand, the empirical data show
deviations of $\Upsilon_k$ from $N$ for the smallest and largest
eigenvalues (except for the largest eigenvalue).  In these cases, the
number of contributing connections is much smaller than $N$ ranging
from a few connections to a few hundreds. These deviations of few
orders of magnitude of $I_k$ from its average suggests that the
vectors are localized \cite{Revmodphys} and that only a few
connections contribute to them. As it will be illustrated on a simple
example in the next section, these results have a clear meaning in the
case of large eigenvalues for which the connections are correlated. In
addition, it was also shown (\cite{Amaral99b} and see below) that
strongly correlated pairs of routers (which correspond to large
components in the eigenvectors) also appear with a relative negative
sign in the eigenvector for {\it small} eigenvalues. This explains why
the lower band edge also displays localized vectors but there is no
clear connection with the spectrum observed in localization in
electronic systems\cite{Revmodphys}.

In addition, our empirical results exhibit `quasi-extended' states in
the center of the band. These states consist essentially of a group of
$\simeq 300-400$ connections corresponding to eigenvalues of order
$0.2-0.4$.

The physical picture which emerges is thus the following. The largest
eigenvalue has an eigenvector which $\Upsilon_{k=1}$ is of order $N$ and
thus represents the whole network. The eigenvectors which correspond
to eigenvalues which deviate from pure random matrix theory correspond
to genuine correlations in the network. We have shown that these
`deviating' eigenvectors (of the order of $20$) have a small value of
$\Upsilon_k$ which means that these important correlations are
localized and that a relatively small number of connections
concentrate most of the activity\cite{Note_flow}.

\subsubsection{Non-Universal Properties: Active Centers}

The detail of the components of the `deviating' eigenvectors give us
information about the important correlations in the network. In
particular, the largest components of the eigenvectors correspond to
the most correlated connections. This can be seen on the simple
following example of a $3\times 3$ correlation matrix
\begin{equation}
\label{matrice}
\left(\begin{array}{ccc}
1 & c & 0\\
c & 1 & c'\\
0  & c' & 1
\end{array}\right)
\end{equation}
where $c$ (resp. $c'$) denotes the strength of the $(1,2)$
(resp. $(2,3)$) correlation. If we denote the ratio of the correlation
strengths $\eta=c'/c$, the eigenvectors $u_1$, $u_2$, and $u_3$ are
respectively
\begin{equation}
\label{e-v}
\left(\begin{array}{c}
1 \\
\sqrt{1+\eta^2}\\
\eta
\end{array}\right),
\left(\begin{array}{c}
-\eta \\
0 \\
1
\end{array}\right),
\left(\begin{array}{c}
1 \\
-\sqrt{1+\eta^2} \\
\eta
\end{array}\right),
\end{equation}
and correspond respectively to the eigenvalues (sorted in decreasing
order)
\begin{equation}
1+c\sqrt{1+\eta^2}\,,1\,,1-c\sqrt{1+\eta^2}
\end{equation}
We thus see on this simple example that the components of the
eigenvector $u_1$ (corresponding to the largest eigenvalue) identify
the most correlated indices: For $\eta\ll 1$, $u_1\simeq (1,1,0)$ and
for $\eta\gg 1$ one obtains $u_1\propto (0,1,1)$.

This remark shows that the eigenvectors are indeed important for
identifying the most correlated connections in the network. We note
that the large correlations are also reflected in the
components---but with a relative minus sign---of the eigenvectors
for small eigenvalues.

In the case of Renater, we have seen in the previous section that all
the components of $u_1$ are positive which indicates a correlation
among the whole network. Even if all the components of $u_1$ indicate
correlations existing in the network, the simple example above shows
that its largest components correspond to the most correlated
connections. We thus looked at the largest components of $u_1$. A
first fact is that a connection $(i,j)$ is always (strongly)
correlated with the connection $(j,i)$. This result is not surprising
since for most operations (Web browsing, Telnet, etc), there is always
a `outgoing' flow which is a significant part of the `incoming' flow.

\begin{figure}
\narrowtext
\centerline{
\epsfysize=0.5\columnwidth{\rotate[r]{\epsfbox{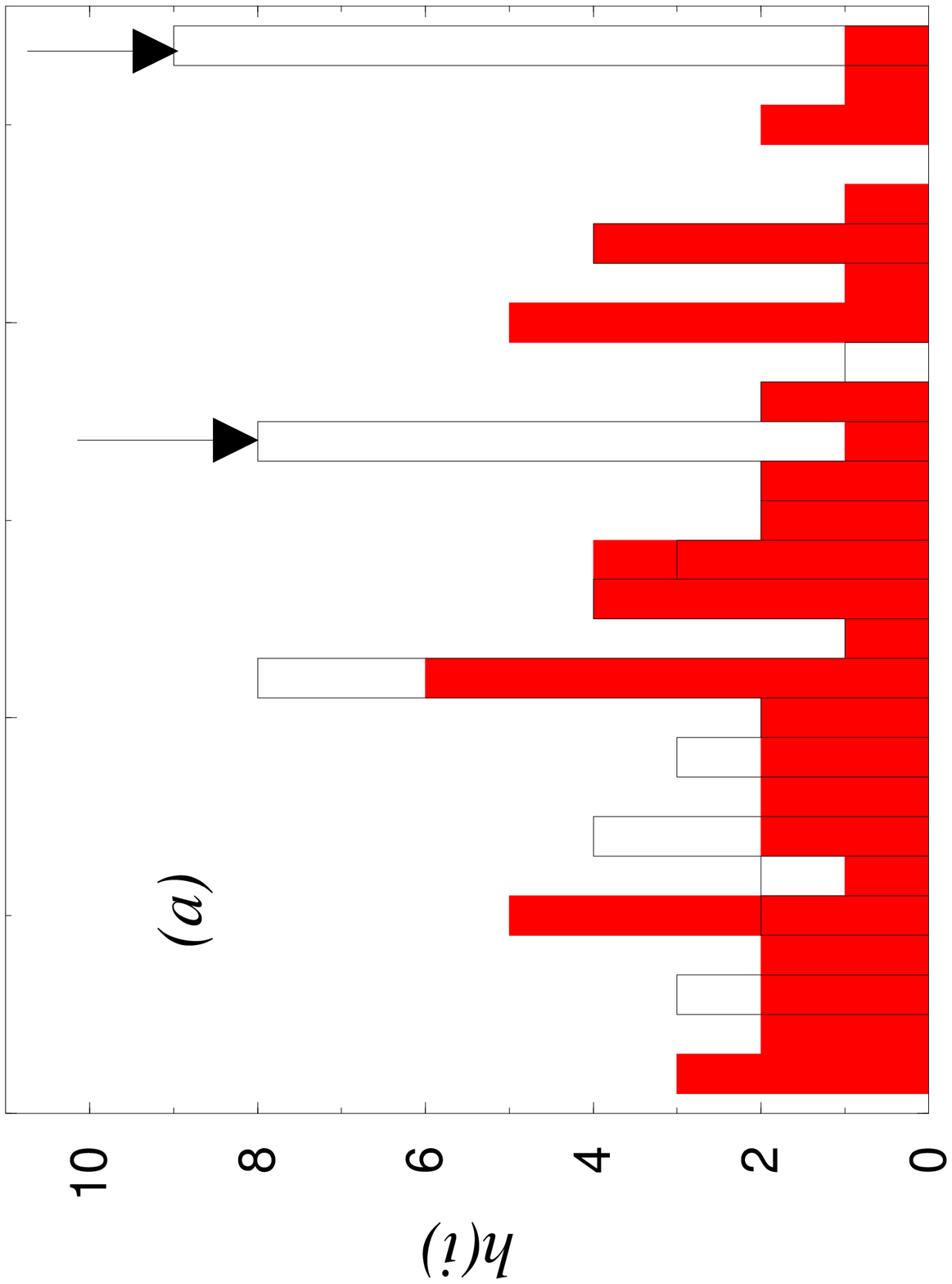}}}
}
\centerline{
\epsfysize=0.5\columnwidth{\rotate[r]{\epsfbox{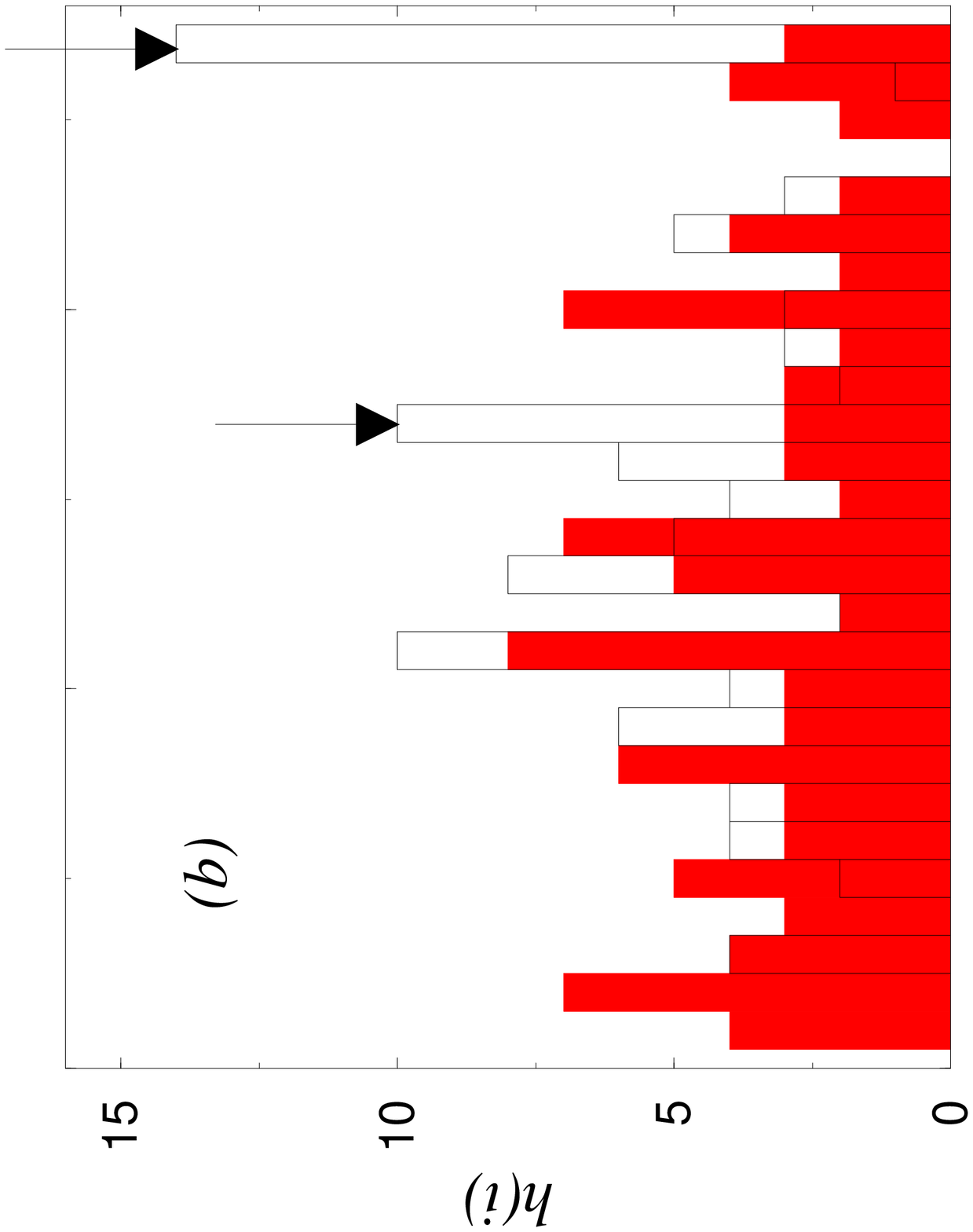}}}
}
\centerline{
\epsfysize=0.5\columnwidth{\rotate[r]{\epsfbox{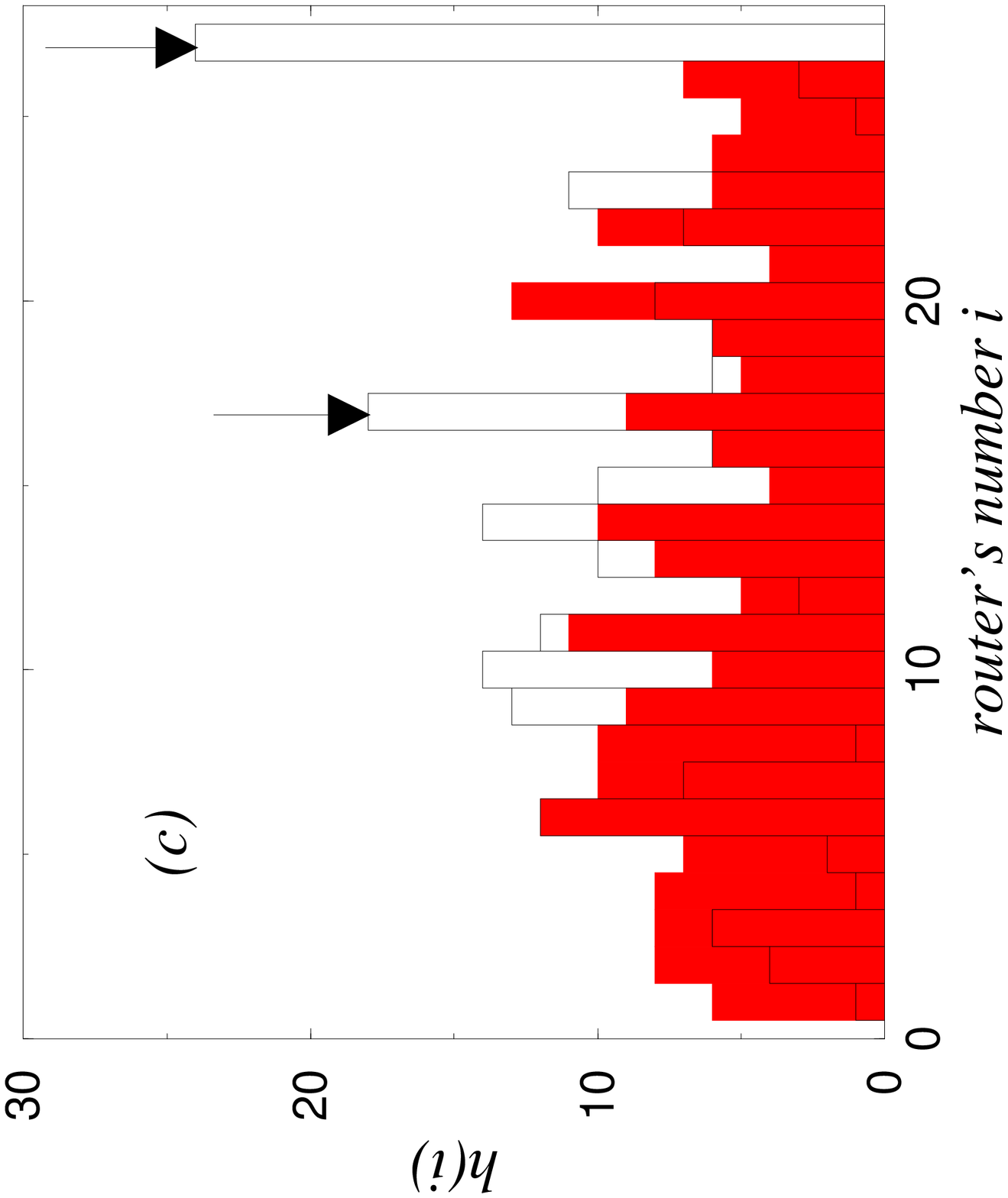}}}
}
\vspace*{0.1cm}
\caption{ Number of occurrences of routers in the $n$ most correlated
connections (There is a total of $26$ routers $i=1,\cdots,27$, the
router $24$ is excluded of the present study for technical
reasons). In each plot, we compared the empirical results with the
control case (histogram in red). The arrows indicate the two most
frequent routers for Renater. In cases (a) $n=30$ and (b) $n=50$, it
is clear that not all routers are participating equally. (c) Case
$n=100$. The control case still fluctuates around its average (which
is $200/26\simeq 7.70$) but much less than the empirical case. This
fact and the observed persistency for increasing $n$ suggest that it
is very unlikely that the empirical peaks are just fluctuations due to
noise. These peaks corresponds probably to routers which are very
active and which are exchanging information with many other routers,
thereby inducing correlations in the network.}
\label{histo}
\end{figure}

In order to look for other causes of correlations we plot on
Fig.~(\ref{histo}) the histogram of occurrences $h(i)$ of the router
$i$ in the set of the $n$ most correlated connections $(i,j)$ which
are given by the first $n$ components of the eigenvector $u_1$
corresponding to the largest eigenvalue. We compared the empirical
results with the control case for increasing values of $n$ (for $n$
approaching the total number of components $N=650$ all the connections
appear and the histogram of occurrences is flat).  We observe marked
differences between these two cases. In particular, in the control
case the histogram tends to be uniform while for Renater we observe
persistent peaks. On the last plot (Fig.~\ref{histo}c), it is apparent
that there are still some fluctuations in the control case but much
less than in the empirical one. The persistency of peaks and the fact
that they appear to be much larger than the average value suggest that
it is very unlikely that they are just fluctuations due to
noise. Therefore, not all routers appear in the most correlated
connections and the peaks can thus be identified as important `active
centers'. These centers are exchanging information with many other
routers thereby inducing correlations between these connections.

It is interesting to note that occurrence peaks also appear in the
components of the other deviating eigenvectors and would thus also
correspond to active centers but at a lower level of correlation.

At this stage, we would like to emphasize that this analysis highlights
active center independently of the volume of information
exchanged. Indeed, in a volume flow analysis the `small' routers even
very active are completely hidden by the `big' routers which are
receiving and emitting huge amounts of bytes.

\section{Correlations and Self-Similarity in the WWW}

The Internet is an example of a complex network which shows existence
of a collective behavior such as a phase transition to a congested
regime\cite{Csabai94}. An important discovery was also the power-law
decay of time correlations\cite{Leland94}. This self-similarity is
usually explained on the basis of underlying distributions of WWW
document sizes, effect of user `think time' and the addition of many
such effects in a network\cite{Crovella97}.

The present study shows that strong correlations between different
connections exist in the traffic network. This result together with
the existence of a phase transition, the existence of a power law
decay of time correlation suggests that the large-scale data traffic
dynamics could be described by a set of simple coupled stochastic
differential equation, such as the Langevin equations with random
interactions\cite{Domi78}. The equation for the Internet activity on a
given connection $(i,j)$ would thus be
\begin{equation}
\frac{\partial g_{ij}}{\partial t}=F(g_{ij}(t))+\varepsilon_{ij}(t)
+\sum_{kl}J_{(ij)(kl)}g_{kl}(t)
\label{spin}
\end{equation}
where the function $F$ is usually expanded for small $g$ as\cite{Golden}
\begin{equation}
F(g)\simeq -rg-ug^3
\end{equation}
and describes the relaxation of a single isolated connection. The
random noise $\varepsilon$ is associated to the effect of users and the
quantity $J_{(ij)(kl)}$ is the coupling between connections $(ij)$ and
$(kl)$.  In the absence of interaction, the correlation function
$<g(t)g(t+\tau)>$ decreases exponentially with a typical correlation
time of order $1/r$ (for $u=0$). When the coupling is strong enough,
the system described by Equ.~(\ref{spin}) undergoes a transition to an
ordered state where all $g$'s are centered around a non-zero value. At
the transition point the correlation function is decaying as a power
law\cite{Golden}.

In this model [Equ. (\ref{spin})], the observed self-similarity in
time is a consequence of the strong correlation existing in the
network. This is in contrast with previous studies which explained the
self-similarity as an effect of existing local power law distribution
(such as the file size distribution). However, more data are needed
for testing this hypothesis and the validity of Equ.~(\ref{spin}) for
the Internet traffic.

\section{Conclusions}

In summary, the largest part of the correlation matrix of connections
is random but also contains statistical information distinct from pure
noise. The eigenvectors which correspond to eigenvalues outside of the
RMT predictions contain information about genuine traffic
correlation. In particular, the largest components of eigenvector
$u_1$ (which corresponds to the largest eigenvalue) indicate the most
correlated connections. We found different origins for the observed
correlations. First, a connection $(i,j)$ is always strongly
correlated with $(j,i)$ which is expected since for each
process---such as web browsing for example---information is exchanged
in both directions. Second, it appears that in the set of the strongly
correlated connections there is only a small number of different
routers which participate in different connections thereby inducing
correlations. This support the idea of the existence of active centers
which are either very active or very visited.  More work and data---on
larger space and time scales---are needed in order to understand more
thoroughly the existence of such centers which seem to play an
important role in the network traffic.

The approach presented in this study thus seems to allow one to
extract relevant correlations between different connections and might
have potential applications to traffic management and optimization. In
particular, this analysis focus on activity independently of the
volume of information exchanged and can thus reveal some very active
routers which are usually hidden by `big' routers exchanging very
large flows.

Finally, the existence of strong correlations together with the
existence of a phase transition and power-law decaying autocorrelation
function suggest that the Internet traffic is similar to a spin glass
close to the critical point. In this hypothesis, the self-similarity
appears naturally as the result of a collective behavior without
resorting to pre-existing power laws.

We thank F.~Baccelli for stimulating and interesting discussions. This
work was supported by the {\it Equipe Reseaux, Savoirs $\&$
Territoires}, Ecole normale Sup\'erieure, Paris.




\end{multicols}


\begin{references}
\bibitem{Note_flow} This localization of {\it activity} has not to be
confused with traffic localization (namely that a subgroup of the
network concentrates most of the traffic volume) which was also
found for this network\cite{Barthelemy02}.
\end{references}
\end{document}